\documentclass[aps,prl,groupedaddress,twocolumn]{revtex4-1}

\setcitestyle{numbers}%

\usepackage[pdftex]{graphicx}
\usepackage{amsmath}
\usepackage{hyperref}
\hypersetup{colorlinks=true, linkcolor=blue, citecolor=blue, filecolor=blue, urlcolor=blue, pdftitle=}
\usepackage{float}
\usepackage{array}
\usepackage{hhline}
\begin{document}


\title{Multimodal Functions as Flow Signatures in Complex Porous Media }


\author{Branko Bijeljic, Ali Q. Raeini, Qingyang Lin, Martin J. Blunt}
\affiliation{Department of Earth Science and Engineering, Imperial College, London SW7 2BP}


\date{\today}

\begin{abstract}

This study refutes the premise that the distribution of flow speeds in complex porous media can be described by a simple function such as a normal or exponential variation. In many complex porous media, including those relevant for subsurface storage and recovery applications, a separation of scales exists between larger inter-granular pores and micro-porosity inside the grains, leading to different flow signatures that need to be described by multimodal functions with distinct flow field characteristics. We demonstrate this finding by devising a novel methodology to simulate fluid flow in carbonate rock based on a representation of the pore space obtained by differential X-ray imaging. The permeability assignment in the micro-porous space is estimated from the pore size inferred from mercury injection porosimetry. Model predictions considering micro-porosity agree well with experimentally measured porosities and permeabilities. The micro-porosity can contribute significantly to the overall permeability, particularly in the more heterogeneous media.

\end{abstract}


\maketitle

The complexity in fluid flow in porous media originates from the presence of pore structures of different size. A striking example are carbonate rocks which account for over 50\% of all hydrocarbon reserves and are attracting widespread interest in applications such as carbon storage and nuclear waste disposal \cite{Bachu2000,Pruess2002,Blunt2013}. Similarly, a large range of pore sizes is encountered in chromatography, packed bed reactors, and fuel cells \cite{Giddings2011,Fogler2016,Mench2008}.  It is therefore difficult to quantify and characterize the flow properties for optimal process design, or to predict recovery.

 An entrenched premise has been that the distribution of velocities in the pore space can be described by a single function with some unique characteristic average value. The observed flow velocity distributions in beadpacks using either NMR imaging or particle tracking velocimetry range from Gaussian \cite {Maier1998}, lognormal \cite{Cenedese1996} , to exponential \cite {Maier1998,Shattuck1991,Lebon1996,Kutsovsky1996,Sederman2001,Moroni2001}. Datta et al. \cite {Datta2013} used confocal microscopy to visualize spatial fluctuations in fluid flow through a dense, disordered beadpack. They found that the velocity magnitudes and the velocity components both along and transverse to the imposed flow direction are exponentially distributed, in agreement with direct numerical simulation on an image of pore space of a beadpack \cite{Bijeljic2013a}. Work on model 2D disordered porous media \cite{Matyka2016} and on stochastically generated 3D porous media \cite{Siena2014} further supported the exponential law as a general form of the velocity distribution functions. Alim et al. \cite{Alim2017} pointed out that the local correlations between adjacent pores, which determine the distribution of flows propagated from one pore downstream, predict the flow distribution. Using numerical simulations of a 2D porous medium they showed the transition of flow distributions from $\delta $-function-like via Gaussian to exponential with increasing disorder. Thus, they demonstrated that a combination of two singular and continuous distributions in flow fractions gives rise to an exponential decay, generalizing results previously obtained in a similar description of force distributions in random bead packs. 


Recent advances in X-ray microtomography have made it possible to describe the geometry of pore structures at micron resolution which has been complemented by the development of direct numerical simulation (DNS) methods on pore-space images to determine flow and transport signatures \cite{Gouze2008,Bijeljic2011a,Bijeljic2011b,Bijeljic2013b,Zaretskiy2010,Gjetvaj2015, Soulaine2016}. However, further experimental characterization of the pore space and its connectivity below micron sizes may be needed to achieve better predictive capabilities of current pore-scale models. To overcome this limitation, differential imaging uses high-salinity contrast brine to non-invasively obtain 3D spatially resolved information on porosity at the sub-micron scale \cite {Lin2016,Lin2017}. Crucially, as the contrast fluid invades the sub-micron pore space, full information of connectivity is provided, which may be a drawback for some other methods such as FIBSEM or SEM, which, while they explicitly resolve the 3D pore structure at high resolution, use small samples, so that the connectivity to the larger pores is not determined. 

{
In this Letter, we develop a methodology that uses X-ray tomography images as a basis to perform direct numerical simulation of flow in both micro- and macro-porous regions. It consists of (a) using the differential imaging method \cite{Lin2016} to obtain a detailed representation of the sub-resolution micro-porous space which complements the voxel-based representation of macro-pores; (b) using mercury injection porosimetry to anchor the assignment of permeability in the micro-porous space; and (c) simulating flow in both micro-porous and macro-porous regions. We study exemplar carbonate rocks shown in Figure 1: a high permeability Ketton limestone (with permeability of   $\mathcal{O}$ $(D)$, 10\textsuperscript{{}-12} m\textsuperscript{2}), an intermediate permeability Estaillades limestone (with permeability of  $\mathcal{O}$ $(100\mathit{mD})$), and a low permeability Portland limestone (with permeability of  $\mathcal{O}$ $(\mathit{mD})$) to quantify the impact of velocity distributions on different macroscopic manifestations of flow. }

\begin{figure}[H]
 \centering 
  \includegraphics[width=0.49\textwidth]{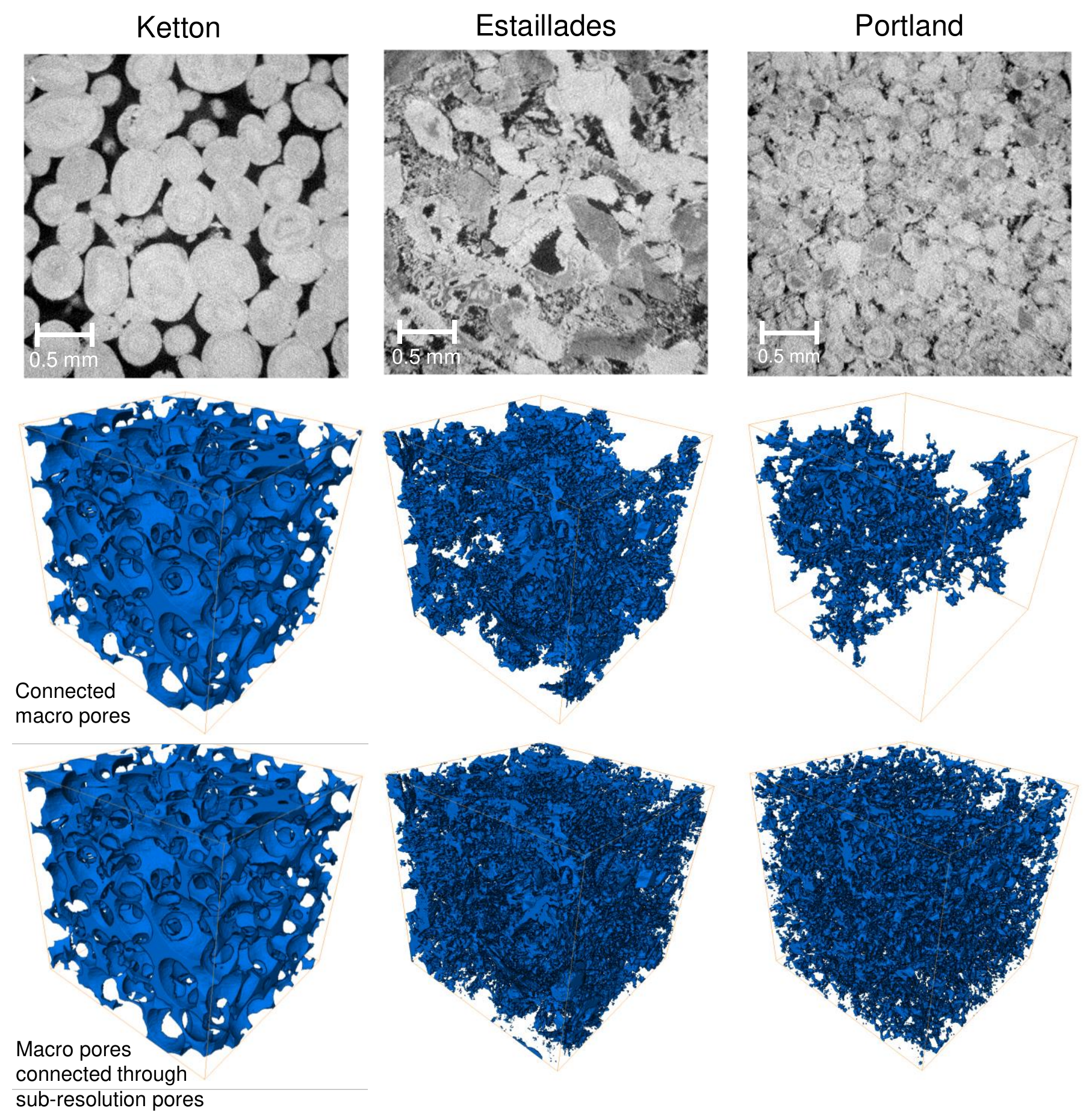} 
\caption{(top) Example two-dimensional slices of a three-dimensional image (with dimensions of  $650^3$  voxels at a voxel size of  $5\mathit{\mu m}$) for the Ketton, Estaillades and Portland samples. (middle) Three-dimensional visualisation of the connected macro pore structure, and (bottom) the macro pore structure connected through sub-resolution pores.}
\end{figure}

{
We will demonstrate that for an accurate description of flow one needs to take into account the separation of scales. That is, families of non-trivial velocity distributions existing in the micro- and macro pore space need to be considered. This requires a more detailed description than possible for a disordered medium with a variation in pore size around some typical value. }



{
$\it{Experimental}$  $\it{ method.-}$  The rock samples used are Ketton, Portland and Estaillades limestones. All three samples were drilled into cylindrical cores 4.81 mm in diameter and 10.0 mm in length. The cores were then placed into a fluoro-polymer elastomer (Viton) sleeve, which was attached to metal fittings (end piece) connecting the core to the pore-fluid flowlines. This assembly was then placed within a Hassler-type flow cell. High pressure syringe pumps were used to maintain pressure and control flow in the pore space. The brine solution was made from deionised water with a prescribed amount of Potassium Iodide ($30\%$ KI). The brine injected from the brine pump was initially pre-equilibrated with the host rock to prevent fluid/solid chemical reaction. After taking the dry (air) scan, the sample was fully saturated with brine (doped with  $30\%$ KI) followed by taking the brine saturated scan. The detailed experimental apparatus configuration and procedure can be found in \cite{Lin2016}.}

{
The imaging was performed using a Versa XRM-500 X-Ray Microscope. The three-dimensional images were reconstructed from a set of 2001 projections. After reconstruction all the images were registered (aligned) according to the reference dry (air) scan -- this was done in order to have the same orientation and position for visualisation and comparison. The images used in later sections are cropped cubic sections with a dimension of  $650^3$  voxels: this represents a bulk volume of  $0.034 cm^3$, or approximately one fifth of the total sample size.}


Image processing method is shown on the raw image for Ketton in \cite{Supplementary2018}. The X-ray differential image method is based on obtaining a difference in grey-scale values between the KI saturated scan and the dry scan. We use the same method for Estaillades and Portland, see \cite{Lin2016}. The high salinity brine absorbs X-rays more strongly than the solid, and hence water-saturated regions appear brighter in the image. The sample was segmented into three phases using a global thresholding method separating grains, sub-resolution pores, and macro pores. Sub-resolution pores are further segmented according to their voxel grey-scale values by the method reported in \cite{Lin2016} which for Ketton identifies two sub-resolution regions with distinct average porosities. The uncertainties caused by the threshold value chosen during image segmentation have also been quantified, as described in \cite{Supplementary2018}. The detailed segmentation method using the Estaillades and Portland samples has been described in more detail in \cite{Lin2016}. The micro-CT image porosity based on segmentation agrees well with Helium porosimetry measurements for all three carbonates\textbf{, }as shown in Table \ref{Table1}.  

\begin{table}[H]
\centering
\small
 \begin{tabular}{| m{1.9cm} |  m{1.9cm}  | m{1.9cm} |  m{1.9cm} |}
\hline
  &  \centering Ketton &  \centering Estaillades &  \centering Portland \arraybackslash \\
 \hline 
 \raggedright  ${\phi  }_{\mathit{tot}}$ from Micro-CT & \centering $ 0.217 \pm 0.008 $  & \centering $ 0.294 \pm 0.014 $   &  \centering   \arraybackslash $ 0.196 \pm 0.011 $  \\\hline
  \raggedright  ${\phi  }_{\mathit{tot}}$ from Helium porosimetry  & \centering  $ 0.234 \pm 0.006$   & \centering  $ 0.293 \pm 0.007 $   &    $ 0.195 \pm 0.006 $ \\\hline
  \raggedright  ${K}_{\mathit{tot}}$ from experimental measurement  & \centering  $ 2.88\pm 0.01  $  $\times {10}^{-12} {m}^{2}$ & \centering   $ 1.75  \pm 0.11 $ $\times {10}^{-13} {m}^{2}$   & \centering  \arraybackslash $ 0.89 \pm 0.19  $  $\times {10}^{-15} {m}^{2} $ \\\hline
  \raggedright  ${K}_{\mathit{tot}}$ from simulation  & \centering  $ 5.06 \pm 0.04  $  $\times {10}^{-12} {m}^{2}$   & \centering  $ 2.06 \pm 0.43  $  $\times {10}^{-13} {m}^{2}$   & \centering \arraybackslash  $ 1.13 \pm 0.26  $  $\times {10}^{-15} {m}^{2}$ \\\hline
  
\end{tabular}
 \caption{Total porosity obtained by differential imaging tomography method is compared to Helium porosity measurements. Permeability obtained from experimental measurements is compared to predictions of the pore-scale numerical simulator based on differential imaging and mercury porosimetry. Both porosity and permeability are compared on the same samples of Estaillades and Portland. The experimental permeability of Ketton was measured at the core scale in \cite{Tanino2012}. }\label{Table1}.
\end{table}

We will also compare predictions for absolute permeability using the methodology developed in this study with the measurements on the same Estaillades and Portland mini-cores. Permeability measurement on the Ketton mini-core was not possible due to very small pressure differences across the sample, so we used a core-scale measurement from \cite{Tanino2012}.

{
To understand the impact of sub-resolution porosity on flow, the regions containing sub-resolution pores for all three samples were subdivided and labelled according to their average porosity, $\phi_i$, shown in Figure 2. The solid grain phase is shown in black and macro pore phase is shown in white. The sub-resolution regions have been partitioned into two phases for Ketton, $k=2$, and into five phases, $k=5$, for Estaillades and Portland.  }

\begin{figure*}[htb]
 \includegraphics[width=0.8\textwidth]{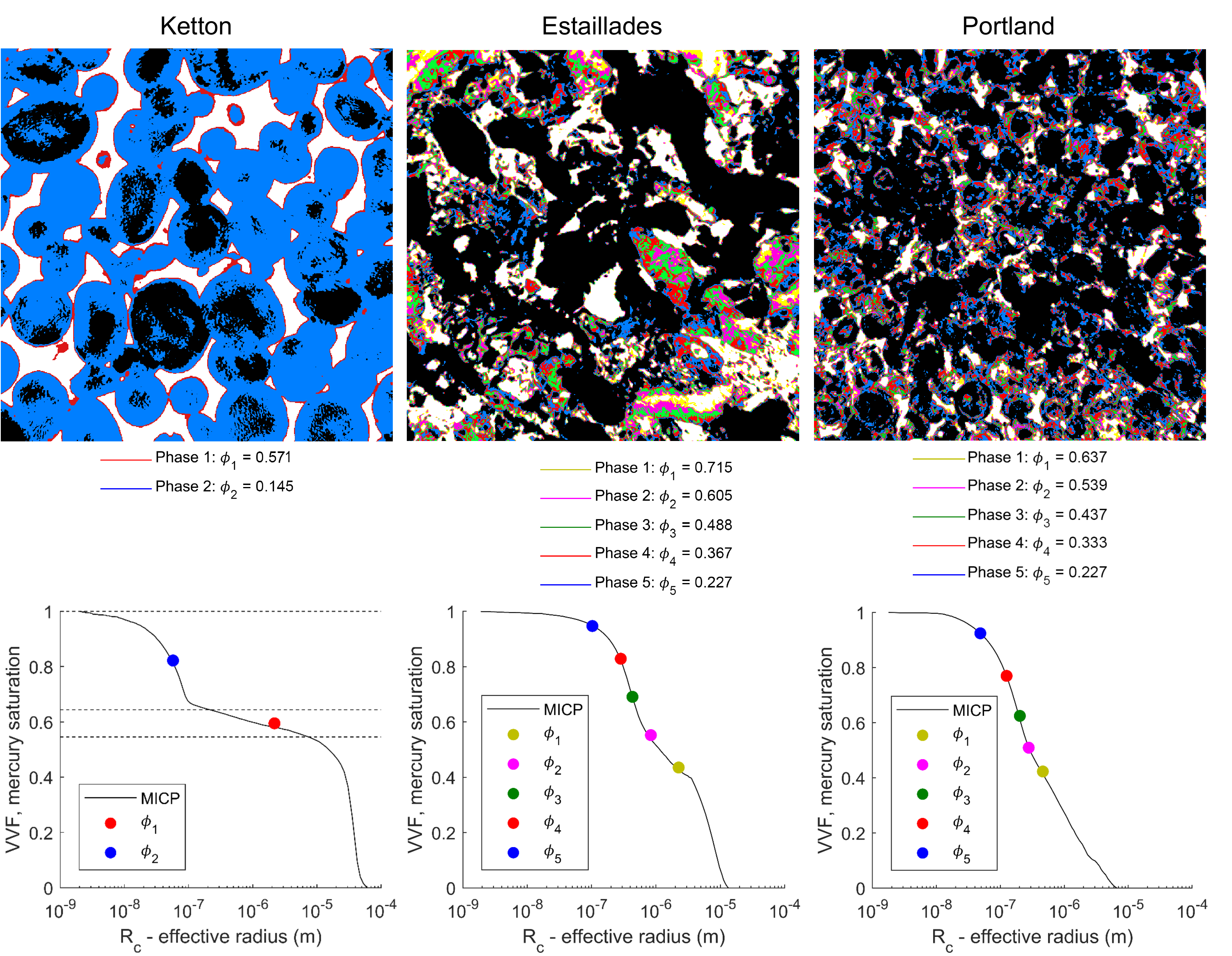} 
\caption{(top) A visualization of void space (black), different phases of the micro-porosity (coloured areas) and the impermeable solid phase (white) for the three carbonate samples. (bottom) A plot of mercury saturation (corresponding to void volume fraction (VVF) as a function of radius obtained from mercury injection capillary pressure, MICP, experiments. The coloured points show the radii assigned to the average value associated with each range of porosity of a micro-porous phase. These points are obtained by assuming the voxels with larger porosity and permeability correspond to the larger radii from the MICP experiments.}
\end{figure*}

To estimate the permeability of sub-resolution phases voxels we first find the effective porosity and pore radius for the micro-porous regions. We estimate the average pore and grain sizes in these voxels by anchoring mercury injection capillary pressure (MICP) data to the void volume fraction (VVF) of each micro-porous phase calculated from the micro-CT image. As illustrated in Figure 2 for the Ketton sample, first we convert the MICP data into effective radii,  $R_c=\frac{2\sigma \cos \theta }{p_c}$, where  $\sigma $  is the interfacial tension of mercury-air,  $485\mathit{mN}/s${, } $\theta $  is its contact angle with the solid phase, assumed to be 45 degrees, and  $p_c$  is the mercury injection capillary pressure measured on the carbonate samples taken from the same block of Ketton, Estaillades and Portland limestone. We assume that the voxels with larger porosity correspond to the larger radii from the MICP experiments and hence have larger permeability.

As illustrated by dotted lines in Figure 2, the lower and upper bounds for the VVF of macro pore space and micro-porosity phases are obtained from the porosity values $\phi_i$. For the upper bounds, $b_i$, we have:

\begin{equation} \label{eq_1}
\mathit{b_i}=\frac{\sum\limits_{j=0}^i \phi_j n_j}{\sum\limits_{j=0}^k \phi_j n_j}
\end{equation}
where the subscript $i=0$ stands for the macro pore space. $i=1,...,k$ represent micro-porosity phases, and $n_j$ is the number of voxels of each phase.

Then the average pore radius for each micro-porous phase is obtained by taking the average of $R_c$ between the lower and upper bounds of mercury saturation ($S$), $b_{i-1}$ and $b_i$,

\begin{equation} \label{eq_2}
 {R}_i = \frac{\int _{b_{i-1}}^{b_i} R_c dS}{{b_i}-b_{i-1}}.
\end{equation}

To assign permeability,  $k_i$, to each micro-porosity voxel, we use the Kozeny-Carman equation \cite{Kozeny1927,Carman1937}, rewritten based on the average pore radius,  $R_i$, assuming that average grain radius is   $d_{g,i}=2R_i\frac{1-\phi _i}{\phi _i}$ :

\begin{equation} \label{eq_2}
k_i=\frac 1{180}\frac{d_{g,i}^2\phi _i^3}{\left(1-\phi _i\right)^2}=\frac 1{45}R_i^2\phi _i
\end{equation}


{
$\it{Mathematical}$  $\it{ Model.-}$ Flow through dual porosity media is modelled using the Stokes equation with an added Darcy term  $\mu k^{-1}\text u$  representing flow in the micro-porous regions which is solved simultaneously with the viscous term  ${\nabla}{\cdot}\left(\mu {\nabla}\text u\right)$, as first introduced by  Brinkman \cite{Brinkman1949}. For slow flow inertial terms can be ignored and the volume averaged momentum and mass conservation equations for incompressible fluids are as follows \cite{Whitaker1986,Ochoatapia1995}:  }

\begin{equation} \label{eq_3}
  \frac{\mu}{\phi}   \nabla^2 \text u =  \nabla p + \mu k^{-1}\text u 
\end{equation}
{

\begin{equation} \label{eq_4}
{\nabla}{\cdot}\text{u=0}
\end{equation}

\noindent where  $u$  is the velocity vector: it is equal to fluid velocity in the void voxels and to the apparent (Darcy) velocity in the micro-porous voxels of the flow domain,   $p$  is the pressure, and  $\rho $  and  $\mu $  are the fluid density and viscosity. Continuity of pressure and velocity is assumed across the boundary region \cite{Ochoatapia1995}. 
For the macro pore voxels, equation \ref{eq_3} simplifies to the standard Stokes equations. For microporous voxels, it leads to the Darcy equation with an additional viscous term which is volume averaged over each voxel that has a porosity $\phi$.

The additional viscous dissipation term in the porous regions ($\mu k^{-1}\text u$) is discretized implicitly: it contributes to the diagonal terms in the matrix of coefficients obtained from the momentum equation. Further details on the discretization and pressure-velocity coupling algorithm are given in \cite{Raeini2012}.  The micro-porous part of the domain has an estimated permeability  $k$  using Eq (3) and the image-measured porosity, while the solid voxels have zero porosity and permeability.  The macro-porous domains are assumed to have  $k={\infty}$ and  $\phi =1$. While similar modelling approaches have been used to describe flow in micro-porosity domains \cite{NorouziApourvari2014,Soulaine2016}, our methodology uses the experimental information from X-ray differential imaging and mercury porosimetry to anchor the porosity and permeability in sub-resolution voxels (Figure 2).

{
The simulations are run at a Reynolds number  $\ Re $ $\ll 1$  until steady-state is reached,  ${\partial}u/{\partial}t=0$. We use constant pressure boundary conditions for pressure at the left and the right faces of the images (the pressure drop is  ${\Delta}P$). The Darcy velocity across the image cross-sectional area is calculated as  $U_D=Q/(L_yL_z), Q[m^3/s]$  is the total volumetric flux calculated as  $Q={\int} u\,dA_x$, where  $A_x[m^2]$  is the cross-sectional area of void voxels perpendicular to the direction of flow  $x$  and  $u$  is the voxel face velocity that is normal to  $A_x$;  $L_x$,  $L_y$,  $L_z$, are the image lengths in each direction. By solving Eq (4) and Eq (5) we obtain the velocities and pressures for each voxel, and calculate absolute permeability  $K[m^2]$  from Darcy's law  $K=(\mathit{\mu Q}L_x)/({\Delta}PL_yL_z)$.}


{
$\it{Results.-}$ Figures 3a and 3b show the distribution functions of velocity magnitudes sampled uniformly in \textit{v},  $P(v)\mathit{dv}$, where  $v = |u| /(\phi U_D)$, is the voxel velocity scaled by the overall Darcy velocity, representing the flow fields in Ketton, Estaillades and Portland on a log-linear and a log-log scale respectively. At first, it appears that it is possible to use an exponential decay function to fit high velocities, which is illustrated in Figure 3 for  $v>10$  for Ketton, corroborating previous studies \cite {Maier1998,Shattuck1991,Lebon1996,Kutsovsky1996,Sederman2001,Moroni2001, Datta2013, Bijeljic2013a, Siena2014, Matyka2016, Alim2017}. The distributions for Estaillades and Portland at first decay faster than for Ketton, but after some threshold exhibit a slower decrease. This is related to the increased heterogeneity in flow field for these lower permeability media with a higher number of both faster and stagnant velocities. }

{
However, the impact of micro-porosity on velocity distributions in complex porous media cannot be captured by a simple exponential decay. In Figure 3c we plot the histograms of the velocity distribution sampled uniformly in bins of  $x=\log _{10}(v)$, $P_l(x) dx$ as introduced in \cite {Bijeljic2011a} which serves to show the low velocity regions of the distributions more clearly. Since  $\int P\left(v\right)\mathit{dv}=\int P_l\mathit{dx}=1$  we have $P_l\left(x\right)=\ln \left(10\right)\mathit{vP}(v)$.}

We can now discriminate between the peaks of high velocities with the values of the order of the Darcy velocity as opposed to the low velocities in the stagnant regions, for all three carbonates studied. In Ketton, a high velocity peak is clearly separated from the stagnant peak due to the remarkable separation of scales between the macroscopic void space and micro-porosity in oolite grains, evident on the micro-CT image and from the pore size distribution inferred from the MICP test (Figure 2). A lower peak of high velocities is observed in the intermediate permeability Estaillades, while the lowest permeability P\bibliographystyle{unsrt}ortland has the smallest number of high velocities. 

\begin{figure}[H]
 \includegraphics[width=0.49\textwidth]{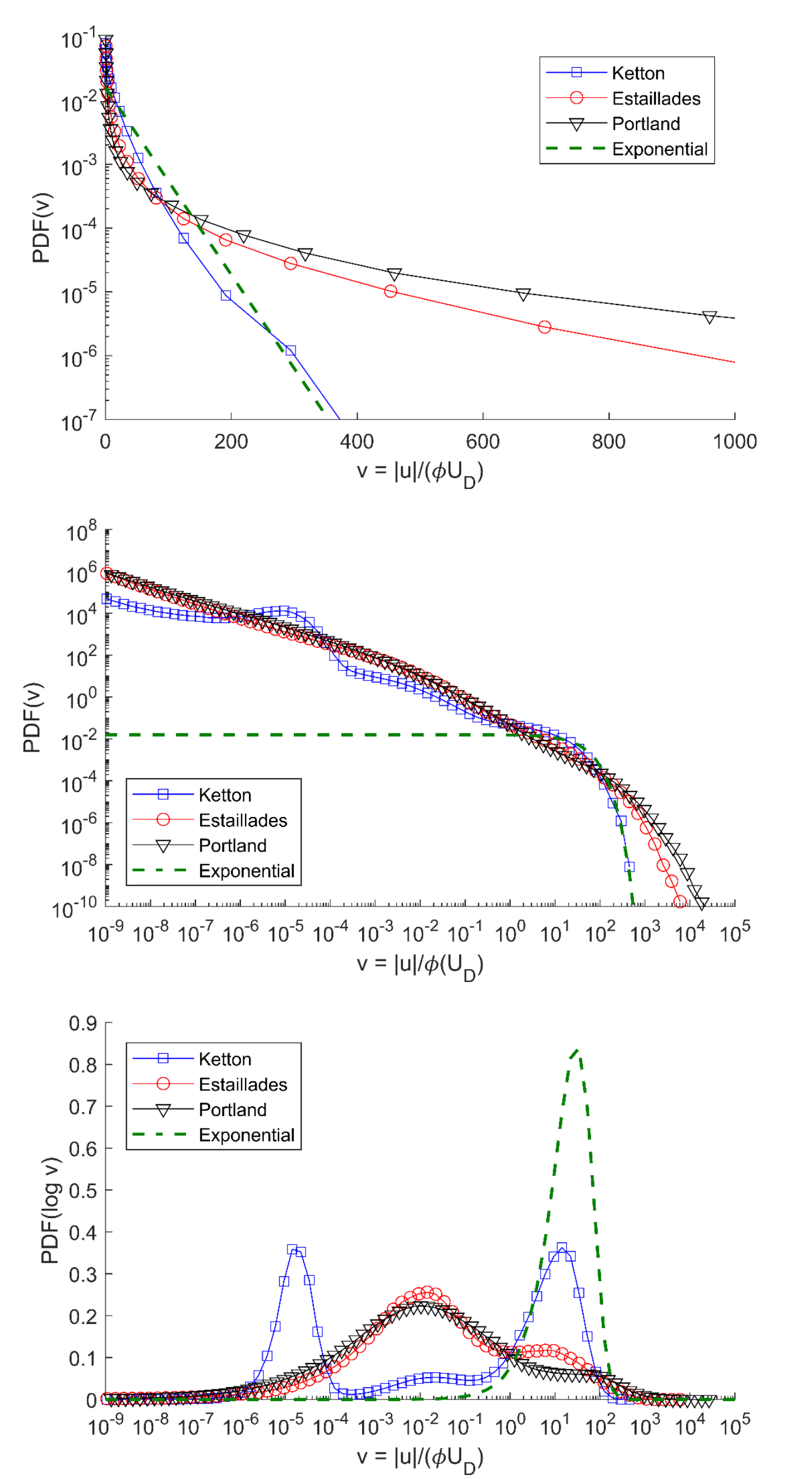} 
 \caption{Probability density functions, PDFs, of voxel velocity magnitude normalized by the Darcy velocity, for the three carbonate rocks on a log-linear scale (top) and on a log-log scale (middle). PDFs of logarithm (base 10) of the normalized velocity (bottom). An exponential decay function is plotted to fit high velocities  $v>10$  for Ketton for comparison.  }
\end{figure}

{
Furthermore, in Figure 4 we plot the distributions of velocities in individual micro-porous phase regions, the sum of all sub-micron phase regions, void macro-pore space phase, and the full velocity distribution for Ketton, Estaillades and Portland. We now discuss the impact of each individual phase on the overall permeability. To discriminate the impact of individual micro-porous phases we run flow simulations to compute the permeability  $K_{c,i}$ {for cumulative porosity}  $\phi_{c,i}${, which starts from considering macro-pore space only, and then adds individual permeable micro-porosity phases in each subsequent simulation starting from the highest micro-porosity region first.} Permeability is presented in Figure 5 as $K_{c,i} / K_{tot} $ versus  $\phi_{c,i}/ \phi_{tot}$  where  $K_{\mathit{tot}}$  is the sample permeability for which all porosities from micro- and macro-pore space are considered.}

\begin{figure}[H]
 \includegraphics[width=0.49\textwidth]{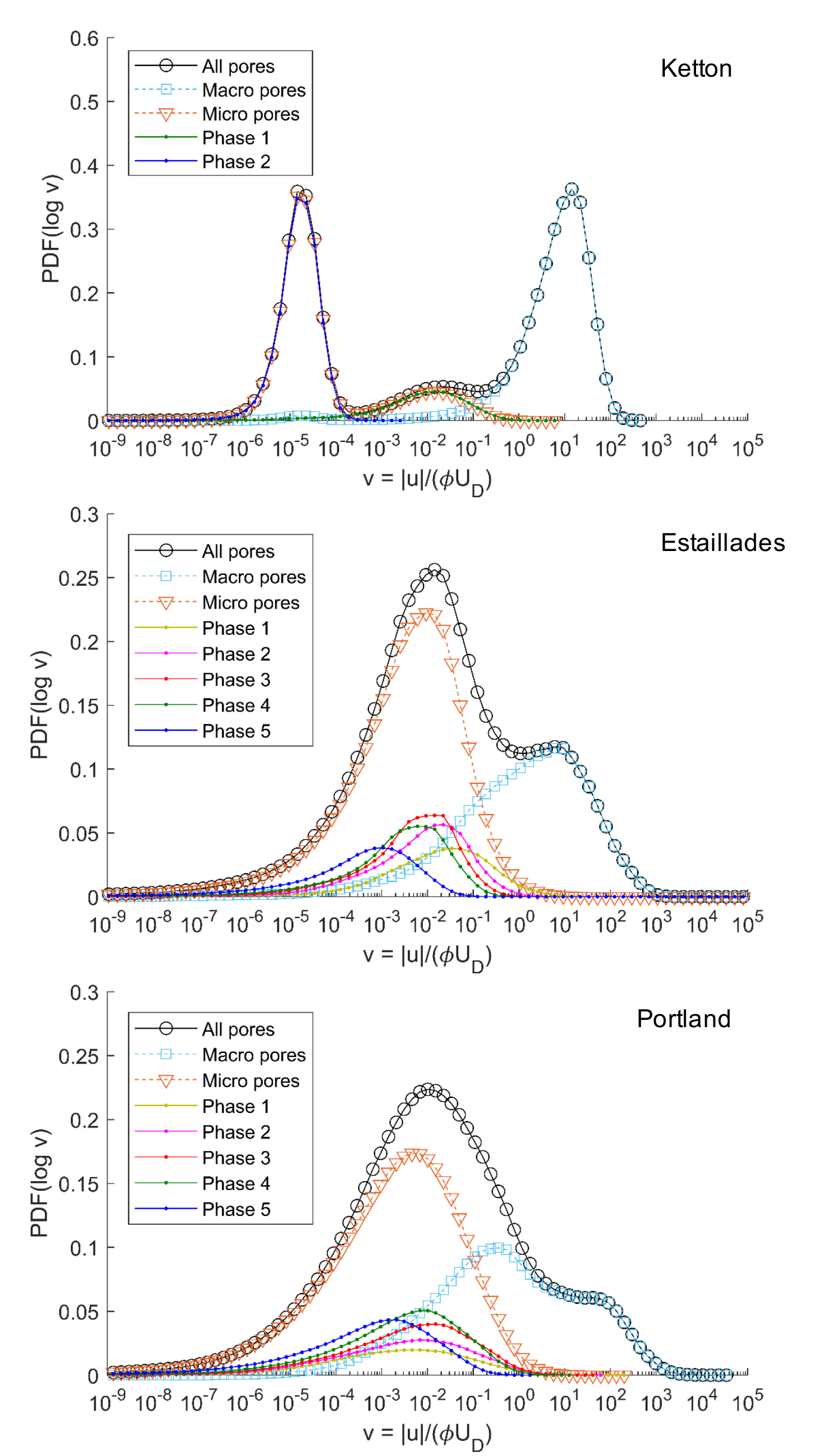} 
 \caption{  PDFs of logarithm (base 10) of the normalized velocity distribution, and the contribution of the velocities from each individual image phase to the overall PDF of the velocity.}
\end{figure}

\begin{figure}[H]
\centering
 \includegraphics[width=0.4\textwidth]{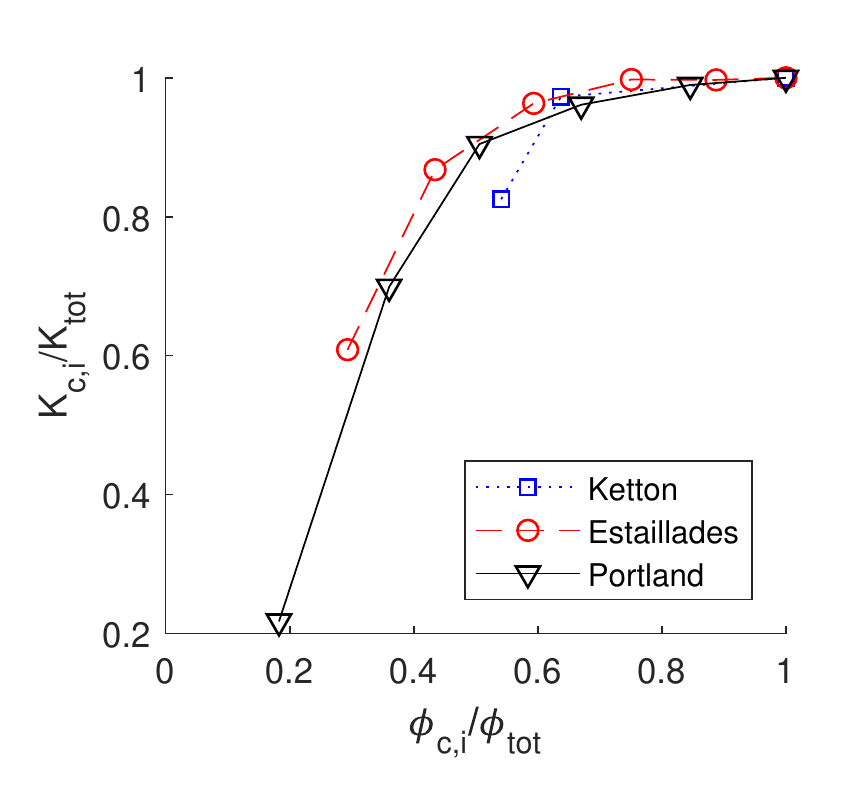} 
 \caption{  Plots of the computed permeabilities of the images as we include different phases of micro-porosity. The left-most point in each plot is the case where no micro-porosity is used in the simulation. The permeabilities are normalized by the permeability of the right-most point in which all micro-porosity phases are included during the simulation.}
\end{figure}

{
Ketton is a remarkable example of contrasting velocities in individual {sub-micron phase regions and void macro-pore space. The low-porosity micro-porous medium has a range of velocities separated from the macro pores, while the higher porosity sub-phase velocities lie between the fast and slow peak and contribute to enhanced flow. The less permeable Estaillades and Portland have a more complex sub-micron pore space in which five phases with different porosity can be observed to have an even higher impact on the permeability increase than for Ketton, see Figure 5. Compared to the permeability of macro-pore space only, when all micro-porosity regions are added this increase is approximately two times for Estaillades and five times for Portland. This indicates that permeability enhancement is more important for the lower permeability porous media. Predictions for permeability based on simulations including all micro-porosity phases compare well with the experimental measurements on the same Estaillades and Portland mini-cores, as shown in  \ref{Table1}.} The uncertainties for permeability values obtained in simulations have been quantified from the values obtained by using the lower and upper bounds of segmented image porosity, as described in \cite{Supplementary2018}. 

{
An important implication of this analysis is that the same velocities in macro-pore and micro-pore space do not have equal weighting on permeability enhancement. While the slow velocities found in the macro-pore space are predominantly near the solid walls of flow channels resulting in flow retardation, the weighting of the slow velocities in micro-porous regions may in fact be such that it enhances flow by connecting otherwise disconnected flow domains in the macro-pore space, see Figure 1. This means that velocity distribution functions from the micro and macro pore space must be considered separately rather than using a single-valued function. }

{
Another implication is that representative elementary volumes based on the macro-pore space are insufficient for the appropriate representation of flow defining permeability. From Figure 2 we observe that sub-resolution phases are non-uniformly distributed thus introducing additional complexity to the representativeness of micro-porous domains when determining volume-averaged properties at a larger scale. }

{
Overall, in complex porous media a separation of scales exists, leading to flow signatures that, instead of displaying a single characteristic velocity scale, need to be described by multimodal functions with distinct flow field characteristics. With the methodology based on X-ray differential imaging, MICP and direct numerical simulation established in this Letter, it is possible to examine and quantify significance of flow in porous media below the micron-scale, which can in turn result in a considerably more complex multiphase flow, transport and reactive transport phenomena including applications in fuel cells, membranes, catalysis and batteries.}

{
B.B. and A.Q.R. would like to acknowledge financial support from Total. }

\end{document}